\documentclass[prb,twocolumn,superscriptaddress,showpacs,floatfix]{revtex4}
\usepackage{amssymb,amsmath}
\usepackage{braket}
\usepackage{graphicx}
\newcommand{\lk}{\left(}
\newcommand{\rk}{\right)}
\newcommand{\lab}{\left|}
\newcommand{\rab}{\right|}
\newcommand{\lkv}{\left[}
\newcommand{\rkv}{\right]}
\newcommand{\lfi}{\left\{}
\newcommand{\rfi}{\right\}}

\newcommand{\bsl}[1]{\begin{slide}{#1}}
\newcommand{\esl}{\end{slide}}
\newcommand{\be}{\begin{equation}}
\newcommand{\ee}{\end{equation}}
\newcommand{\ben}{\begin{enumerate}}
\newcommand{\een}{\end{enumerate}}
\newcommand{\bit}{\begin{itemize}}
\newcommand{\eit}{\end{itemize}}
\newcommand{\been}{\begin{displaymath}}
\newcommand{\eeen}{\end{displaymath}}
\newcommand{\ba}{\left[\begin{array}}
\newcommand{\ea}{\end{array}\right]}
\newcommand{\bac}{\begin{array}}
\newcommand{\eac}{\end{array}}
\newcommand{\bc}{\begin{center}}
\newcommand{\ec}{\end{center}}
\newcommand{\bea}{\begin{eqnarray}}
\newcommand{\eea}{\end{eqnarray}}
\newcommand{\bean}{\begin{eqnarray*}}
\newcommand{\eean}{\end{eqnarray*}}
\newcommand{\bqu}{\begin{quote}\begin{it}}
\newcommand{\equ}{\end{it}\end{quote}}

\newcommand{\hA}{\hat{A}}
\newcommand{\hB}{\hat{B}}

\newcommand{\vk}{\mathbf{k}}

\newcommand{\vvr}{\mathbf{r}}

\newcommand{\vvR}{\mathbf{R}}

\newcommand{\vT}{\mathbf{T}}

\DeclareMathOperator{\Tr}{Tr}
\DeclareMathOperator{\imag}{Im}
\DeclareMathOperator{\rere}{Re}
\begin{document}
\bibliographystyle{apsrev}
\title{Theory of quasiparticle spectra for Fe, Co, and Ni: bulk and surface}
\author{Alexei Grechnev}
\affiliation{Department of Physics, University of Nijmegen, NL-6525 ED Nijmegen, The Netherlands}
\author{I. Di Marco}
\email{dimarco@science.ru.nl}
\affiliation{Department of Physics, University of Nijmegen, NL-6525 ED Nijmegen, The Netherlands}
\author{M. I. Katsnelson}
\affiliation{Department of Physics, University of Nijmegen, NL-6525 ED Nijmegen, The Netherlands}
\author{A. I. Lichtenstein}
\affiliation{Institute of Theoretical Physics, University of Hamburg, 20355 Hamburg, Germany}
\author{John Wills}
\affiliation{Theoretical Division, Los Alamos National Laboratory, Po Box 1663, Los Alamos, NM87545 Usa}
\author{Olle Eriksson}
\affiliation{Department of Physics, Uppsala University, Box 530, SE-751 21 Uppsala, Sweden}

\date{\today}

\begin{abstract}
The correlated electronic structure of iron, cobalt and nickel is
investigated within the dynamical mean-field theory 
formalism, using the newly developed full-potential LMTO-based
LDA+DMFT code. Detailed analysis of the calculated electron
self-energy, density of states and the spectral density are
presented for these metals. It has been found that all these
elements show strong correlation effects for majority spin
electrons, such as strong damping of quasiparticles and formation
of a density of states satellite at about $-7$ eV below the Fermi
level. The LDA+DMFT data for fcc nickel and cobalt (111) surfaces
and bcc iron (001) surface is also presented. The electron self
energy is found to depend strongly on the number of nearest
neighbors, and it practically reaches the bulk value already in
the second layer from the surface. The dependence of correlation
effects on the dimensionality of the problem is also discussed.
\end{abstract}
\maketitle

\section*{Introduction}
The late 3d metals iron, cobalt, and nickel and their compounds
are vital for nearly all fields of technology. The Earth core is
believed to be composed predominantly of iron. It is ironic that
in the early 21-st century we still lack a complete understanding of
these metals. Properties of ``normal'' (weakly correlated) solids
are described quantitatively by density functional theory
(DFT)\cite{hohenberg64pr136:B864,kohn65pr140:A1133} in the local
density (LDA) or generalized gradient (GGA) approximations. Not
only DFT provides the ground state properties, but in many cases
it also gives a rather good description of excitation spectra in
terms of Kohn-Sham quasiparticles. The concept of
quasiparticles originates from Landau's Fermi liquid theory and
for weakly correlated solids the quasiparticles (electrons and
holes) are well defined in a wide energy range.

Fe, Co and Ni, however, are more correlated systems. They have
partially filled shells of fairly localized 3d electrons. These
electrons form a narrow d-band, and their behavior shows signs of
both atomic-like and free-electron-like
behavior\cite{vonsovsky93pmm76:247,lichtenstein01prl87:067205}. In
strongly correlated systems the quasiparticle picture
breaks down, except in a close vicinity of the Fermi surface.
Quasiparticles often have short lifetimes and therefore are not
well defined, and in many cases incoherent features such as
Hubbard bands and satellites appear in excitation spectra\cite{georges96rmp68:13,kotliar06rmp78:865}
. LDA and GGA typically fails for this class of
systems and their theoretical description remains a great
fundamental challenge. In particular, LDA and GGA give rather good
magnetic moments for Fe, Co, and Ni, but fail to describe their
electronic structure adequately. In particular, photoemission
experiments for these
metals\cite{chanderis83prb27:2630,gutierez97prb56:1111,guillot77prl39:1632}
demonstrate that LDA/GGA calculations give too wide majority spin
3d band, overestimate the spin splitting and fail to reproduce the 6
eV satellite in nickel, an essentially incoherent feature. Some
other theoretical methods are needed to properly describe the
electronic structure of Fe, Co and Ni.

One of the successful schemes for correlated electron systems is the
Dynamical Mean-Field Theory (DMFT), which replaces the lattice problem
with a problem of a single correlated site in a self-consistent bath (impurity
problem). It has been originally developed for the Hubbard
model\cite{metzner89prl62:324,georges92prb45:6479,georges96rmp68:13}.
Being $W$ the bandwidth and $U$ the Coulomb interaction,
the DMFT catches the main features of weakly ($W \gg U$), intermediate
($W \sim U$) and strongly ($W \ll U$) correlated regimes, and 
becomes exact in the limit of infinite dimensions. The crucial point of the DMFT
 is in the solution of the self-consistent impurity problem. The
choice of the DMFT ``solver'' for a given system is always a compromise
between generality, accuracy and efficiency.  There exist both
numerically exact solvers (quantum Monte-Carlo, exact
diagonalization) which can in principle be applied to all systems,
and approximate solvers with limited area of applicability but
high efficiency, such as the Spin Polarized T-matrix Fluctuation exchange
(SPTF) solver\cite{katsnelson02epjb30:9} for the case $W \lesssim U$.

Although DMFT was designed originally for the Hubbard model, it
can be combined with LDA to describe realistic materials with a local
electron correlation. This approach, known as
LDA+DMFT\cite{lichtenstein98prb57:6884,anisimov97jpcm9:7359,kotliar06rmp78:865}
is at present the most universal practical
technique for calculating the electronic structure of strongly correlated solids. LDA+DMFT has
been successfully applied to various important problems,
including, e.g., the electronic structure of
manganese\cite{biermann04zhetfl80:714},
$\delta$-Pu\cite{savrasov01nature410:793,pourovskii06epl74:479}, the $\alpha$--$\gamma$
transition in cerium\cite{held01prl87:276404} and the
metal-insulator transition in V$_2$O$_3$\cite{held01prl86:5345}.
Despite all success stories of LDA+DMFT, the method is still less than a
decade old and at a stage of active development. Most available
implementations apply some drastic simplifications to the LDA+DMFT
formalism. In particular, many LDA+DMFT codes are based on atomic
sphere approximation (ASA)-based LDA codes (LMTO-ASA\cite{lichtenstein98prb57:6884}
 or KKR-ASA\cite{minar05prb72:045125}).
These schemes might work well for close-packed crystal structures, but
they are insufficient for open structures and low-dimensional
geometries. However, until recently, the only all-electron
full-potential LDA+DMFT implementation has been the one based on
the full-potential LMTO code LMTART of
Savrasov\cite{savrasov01nature410:793}.

The goal of this paper is to investigate the correlated electronic
structure of bulk and surface of iron, cobalt and nickel using 
the new full-potential LDA+DMFT code BRIANNA. Although iron and nickel have been previously
investigated using ASA-based LDA+DMFT
codes\cite{katsnelson99jpcm11:1037,lichtenstein01prl87:067205,katsnelson02epjb30:9},
no full-potential results have been published. The LDA+DMFT
spectral density of nickel has never been published, while for
cobalt only three-body scattering
approximation\cite{monastra02prl88:236402} data are available, but
no LDA+DMFT results. Further, and most importantly, LDA+DMFT methods
have not been previously applied to transition metal surfaces.
Now, with the present full-potential LDA+DMFT scheme available, we want to check
how the correlation effects depend on the dimensionality of the
problem. We address fcc nickel  and cobalt (111) surfaces and iron
(001) surface in this paper, as examples.

This paper is organized as follows. In chapter \ref{ch:lpd} we
present the LDA+DMFT formalism in its most general form following
a discussion of the basis set problem in
Refs.\onlinecite{anisimov05prb71:125119,savrasov04prb69:245101}.
Chapter \ref{ch:brianna} introduces our full-potential LDA+DMFT
implementation BRIANNA. Results of our calculations are presented
in chapter \ref{ch:results}, which is followed by the conclusion.

\begin{figure}
\includegraphics[scale=0.3]{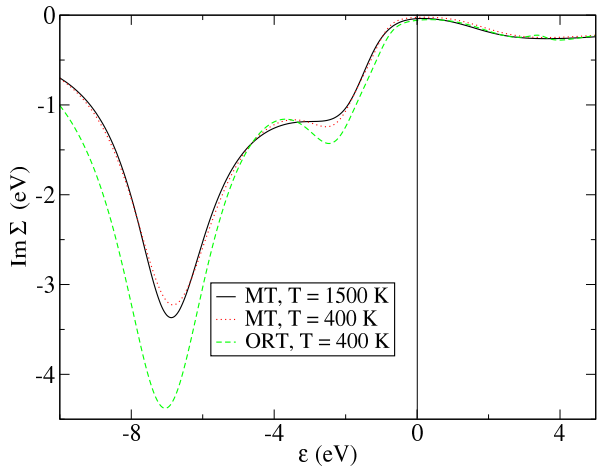}
\caption{\label{f:fe-sig-up} (Color online) Imaginary part of the self energy $\imag\Sigma(\epsilon + i 0)$ for bcc iron, majority spin for
orthogonalized LMTO (ORT) and muffin-tin-only (MT) correlated subspaces. Two different temperatures are
considered: $T=400$ K and $T=1500$ K.}
\end{figure}

\begin{figure}
\includegraphics[scale=0.3]{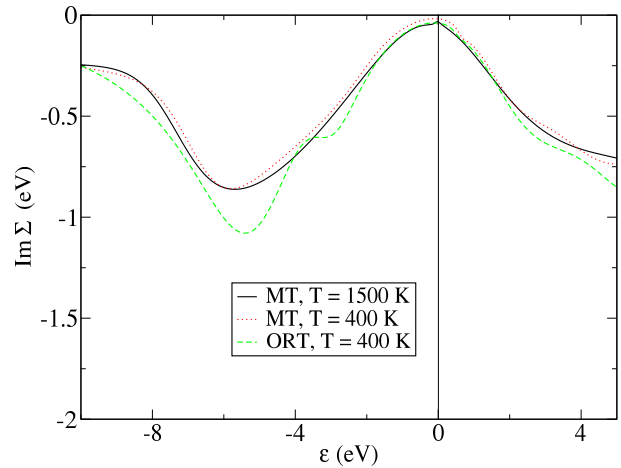}
\caption{\label{f:fe-sig-dn} (Color online) Imaginary part of the self energy for bcc iron, minority spin for two
different correlated subspace(MT and ORT orbitals) and two different temperatures ($T=400$ K and $T=1500$ K).}
\end{figure}

\begin{figure}
\includegraphics[scale=0.3]{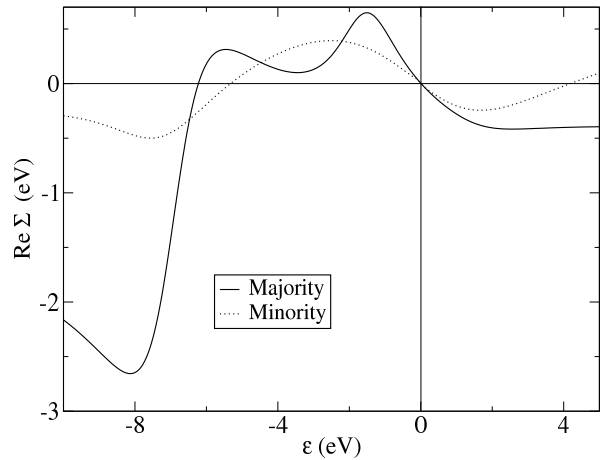}
\caption{\label{f:fe-sig-re} Real part of the self energy for bcc iron  (MT orbitals, $T=1500$ K).}
\end{figure}

\section{\label{ch:lpd}LDA+DMFT method}
\subsection{Correlated subspace}

The LDA+DMFT method defines a "correlated subspace" $\lfi \ket{\vvR, \xi} \rfi$
of the strongly correlated orbitals
$\ket{\vvR, \xi}$, where $\vvR$ stands for the Bravais lattice site and the quantum number
$\xi$ specifies the correlated orbitals within the unit cell. Within this subspace the many-body problem
is solved in a non-perturbative manner using DMFT. All remaining states of the crystal are
assumed weakly correlated and treated within LDA. For simplicity, we can always choose correlated orbitals
to be orthogonal and normalized
$\braket{ \vvR_1, \xi_1 | \vvR_2, \xi_2} = \delta_{\vvR_1, \vvR_2} \delta_{\xi_1, \xi_2}$.

\begin{figure}
\includegraphics[scale=0.3]{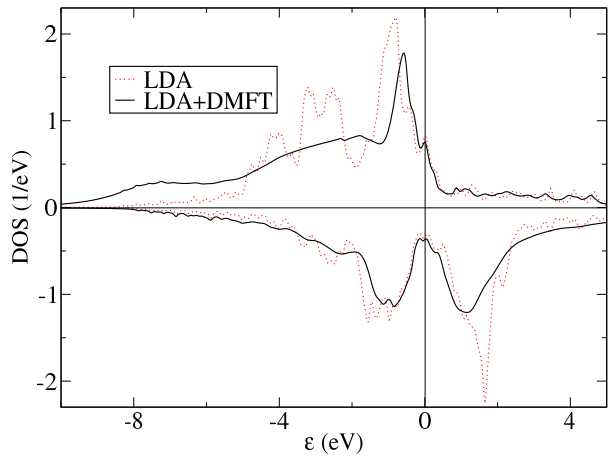}
\caption{\label{f:fe-dos} (Color online) Density of states of bcc iron (LDA+DMFT with MT correlated orbitals at
$T=400$ K vs LDA).}
\end{figure}

\begin{figure*}
\includegraphics[scale=0.8]{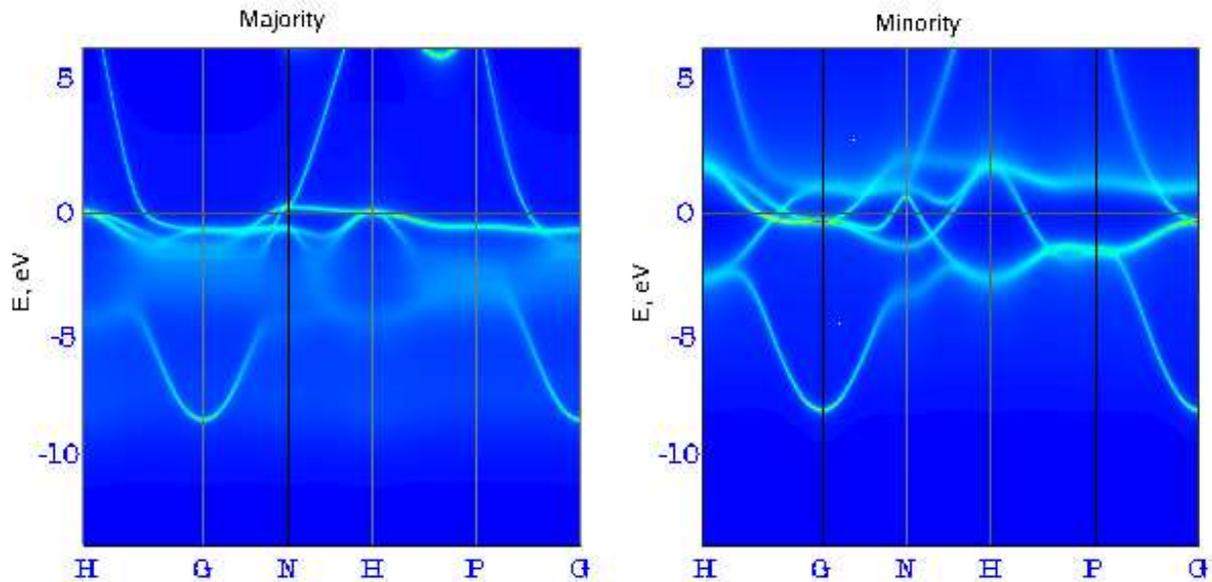}
\caption{\label{f:fe-spec} (Color online) Spectral density of bcc iron along high simmetry directions of the Brillouin
zone from LDA+DMFT with MT correlated orbitals at
$T=400$ K.}
\end{figure*}

Results of a LDA+DMFT calculation depend on the choice of the
correlated orbitals. The correct form of $\ket{\vvR,\xi}$ is
dictated by physical considerations for each particular
problem\cite{anisimov05prb71:125119,savrasov04prb69:245101,lecher}.
Usually the correlated orbitals are derived from $d$ or $f$ atomic
states. In this case $\xi$ stands for site (within unit cell) and
atomic quantum numbers $l,m,\sigma$. In early LDA+DMFT
implementations $\ket{\vvR, \xi}$'s were taken as orthogonalized
muffin tin orbitals (MTO's). Apart from the technical simplicity,
however, there is no reason for making such a choice. In
particular, orthogonalized MTO's are poorly localized in real
space (due to orthogonalization) and they don't have pure $lm$
character (due to tail cancellation and orthogonalization).
Besides, such choice of $\lfi \ket{\vvR,\xi} \rfi$ de facto
restricts the LDA part of the LDA+DMFT code to muffin-tin based
methods (LMTO, NMTO, KKR) with a minimal basis set. Nowadays other
choices, such as Wannier-like
orbitals\cite{anisimov05prb71:125119,lecher} are investigated.
Apart from atomic states, the correlated orbitals $\ket{\vvR,
\xi}$ can be chosen as hybridized orbitals (possibly describing
covalent bonds) if dictated so by the physical problem. Sometimes
the LDA Hamiltonian is downfolded in order to include only
correlated degrees of freedom (such as d-states or even $t_{2g}$
or $e_g$ states). This results in a very time efficient DMFT
implementation, however this approach cannot be used to study the hybridization between
sp-electrons and the correlated ones, which is important, e.g., for
the superexchange interaction. In this section, we present
the LDA+DMFT formalism in a more general form, without making any
restrictions on the choice of the correlated subspace $\lfi
\ket{\vvR,\xi} \rfi$ or the basis set used by the LDA part of the
code. We are going to return, however, to the question of choosing
the correlated subspace in the section describing our
implementation.

In LDA a solid is described by the one-particle Kohn-Sham equation
\be
\label{ks}
\lk H_{LDA} - E \rk \ket{\psi} = 0,
\ee
where the LDA Hamiltonian $H_{LDA}$ has one-electron form
\be
H_{LDA} = \sum_i h_{LDA}(\vvr_i),
\ee
with $h_{LDA}$ acting in
the Hilbert space of one-electron states in a periodic crystal,
and the index $i$ numbering all electrons in the crystal.
The LDA+U Hamiltonian adds explicit Coulomb term for the correlated orbitals
$\ket{\vvR, \xi}$ to the LDA Hamiltonian
\begin{widetext}
\be
\label{hlpu}
H_{LDA+U} = H_{LDA} + \frac 12 \sum_{\vvR} \sum_{\xi_1,\xi_2,\xi_3,\xi_4} U_{\xi_1,\xi_2,\xi_3,\xi_4}
c^{\dagger}_{\vvR,\xi_1} c^{\dagger}_{\vvR,\xi_2} c_{\vvR,\xi_4} c_{\vvR,\xi_3}.
\ee
\end{widetext}
It has a form of a multiband Hubbard Hamiltonian with the LDA
Hamiltonian used as "hopping". The Coulomb parameters
$U_{\xi_1,\xi_2,\xi_3,\xi_4}$ are screened Coulomb integrals for
the states $\lfi \ket{\vvR, \xi} \rfi$. They are to be found
empirically or calculated from first principles, and they, in
general, depend on the choice of the correlated subspace $\lfi
\ket{\vvR, \xi} \rfi$. Note that $H_{LDA}$ is not a kinetic energy
and the Hubbard-U term is not a "raw" Coulomb interaction. Rather,
the LDA+U Hamiltonian is an effective Hubbard Hamiltonian for the
correlated orbitals, and the rest of the states (e.g. sp states) is described
within LDA. The justification of this
approach\cite{savrasov04prb69:245101} is not a trivial matter, and
the main practical reason why it is widely used is the local nature
of the screened Coulomb interaction.
Note that we have included the explicit Coulomb term in
the Hamiltonian, although many static effects of the Coulomb interaction
are already included in the LDA Hamiltonian. Namely, LDA includes
a Hartree term, an exchange term, and some correlation effects
(including a good description of the screening). However, the many-body
treatment of the Hubbard-U part of the Hamiltonian will
also give the Hartree-Fock term and various correlation terms.
Therefore, a double-counting correction scheme is necessary
\be
\label{dcount} \Sigma(z) \rightarrow \Delta\Sigma(z) \equiv
\Sigma(z) - \Sigma_{dc}, \ee
where $\Sigma(z)$ is the local self energy
of the DMFT problem. For treating metals within LDA+DMFT
the most common choice of the double-counting correction is the static part of the self-energy:
\be
\Sigma_{dc} = \Sigma(+i0).
\ee

The multiband Hubbard Hamiltonian, such as Eq. \ref{hlpu} cannot be solved exactly, therefore
various approximations are applied. Unlike the static LDA+U method\cite{anisimov91prb44:943},
LDA+DMFT takes into account dynamical corelation effects through the frequency-dependent
self energy $\Sigma(z)$.

\subsection{LDA+DMFT equations}

Spectral density functional theory\cite{savrasov04prb69:245101,kotliar06rmp78:865,chitra01prb63:115110}
uses the local Green function $G_{\vvR}(z)$ as the main observable quantity, much like the particle density
$\rho(\vvr)$ in DFT or the one-electron Green function $G(z)$ in Baym-Kadanoff theory. Namely, for
sensible definitions of $G_{\vvR}(z)$, a functional $\Gamma[G_{\vvR}]$ exists, which
is minimized by the true value of $G_{\vvR}(z)$, and the local self energy $\Sigma_{\vvR}(z)$
plays the role similar to the Kohn-Sham potential in DFT. In the present paper we define $G_{\vvR}(z)$ as
the projection of the total one-electron GF to the correlated states $\ket{\vvR,\xi}$ of a given site $\vvR$
\be
\label{gloc}
 G_{\vvR}(z) =  P_{\vvR}  G(z)  P_{\vvR},
\ee
where
\be
 P_{\vvR} = \sum_{\xi} \ket{\vvR,\xi}\bra{\vvR,\xi}
\ee
is the projection operator to the correlated subspace belonging to site $\vvR$.
The one-electron GF $ G(z)$ can in turn be expressed via the one-electron
self energy $\Sigma(z)$ as
\be
\label{gf}
 G (z) = \lkv  (z-\mu) -  h_{LDA} -  \Sigma(z) \rkv^{-1},
\ee
where $\mu$ is the chemical potential, and $h_{LDA}$ plays the role of the unperturbed Hamiltonian
("hopping").

Precisely like in DFT or Baym-Kadanoff theory, the exact expression for the functional $\Gamma[ G_{\vvR}]$
is not known. The most widely used approximation is the dynamical mean field theory (DMFT). The approximation behind
DMFT is that the total one-electron self energy $ \Sigma(z)$ is taken as the sum of the local self energies of all
lattice sites (with double-counting correction when appropriate)
\be
\label{dmftsigma}
\Sigma(z) = \sum_{\vvR} \Sigma_{\vvR}(z).
\ee
This self energy is local, i.e. it does not have matrix elements between different sites
\be
\braket{\vvR_1,\xi_1| \Sigma | \vvR_2,\xi_2} = \delta_{\vvR_1, \vvR_2} \lk \Sigma_{\vvR} \rk_{\xi_1,\xi_2}.
\ee
With this form of $ \Sigma(z)$, the local self-energy $ \Sigma_{\vvR}(z)$ can
be obtained from the impurity problem, with the rest of the lattice replaced by the bath GF
(or "dynamical mean field") ${\mathcal{G}}_0^{-1}(\vvR,z)$ defined by
\be
\label{g0def}
{\mathcal{G}}_0^{-1}(\vvR,z) =  G_{\vvR}^{-1}(z) + \Sigma_{\vvR}(z).
\ee
In a periodic solid all atoms are equivalent, therefore the local quantities such
as $\lk \Sigma_{\vvR} \rk_{\xi_1,\xi_2}$,  $\lk G_{\vvR} \rk_{\xi_1,\xi_2}$ and
$\lk {\mathcal{G}}_0^{-1}(\vvR,z) \rk_{\xi_1,\xi_2}$ do not depend on the site
$\vvR$. Moreover, if there are several sites within unit cell, and the Hubbard-U term
does not have matrix elements between different sites,  $\lk \Sigma_{\vvR} \rk_{\xi_1,\xi_2}$
takes a block diagonal form with a block for each site.

The system of one correlated site in the self-consistent bath does not have a Hamiltonian, but can be described by an 
effective action.
Here and in the following we use the Matsubara formalism for
a finite temperature $T$. The GF's and self energies are defined
at the Matsubara frequencies
\be
\label{omegan}
z = i \omega_n = i \pi T (2n+1) \;,\quad n= 0, \pm 1, \pm 2, \ldots \;.
\ee
The action in the Matsubara formalism is
\begin{widetext}
\be
\label{action}
S = - \iint dx_1 dx_2
 c^{\dagger}(x_1) \mathcal{G}^{-1}_0(x_1,x_2) c(x_2) +
\frac 12 \int dx_1 dx_2 dx_3 dx_4  c^{\dagger}(x_1) c^{\dagger}(x_2) U(x_1,x_2,x_3,x_4) c(x_4) c(x_3),
\ee
\end{widetext}
where $x \equiv (\xi, \tau)$, $\tau$ is the imaginary time ($0 < \tau < 1/T$) and
\be
\mathcal{G}^{-1}_0(x_1,x_2) = T \sum_{i\omega_n} \exp\lkv - i \omega_n (\tau_1 - \tau_2) \rkv
 \lk \mathcal{G}^{-1}_0 \rk_{\xi_1,\xi_2}.
\ee
The Coulomb interaction $U(x_1,x_2,x_3,x_4)$ does not depend on time for the Hubbard model
\be
U(x_1,x_2,x_3,x_4) = \delta(\tau_1 - \tau_2)  \delta(\tau_1 - \tau_3) \delta(\tau_1 - \tau_4) U_{\xi_1,\xi_2,\xi_3,\xi_4}.
\ee

This problem still cannot be solved exactly, however, compared to the original many-body problem
it has only a few degrees of freedom, so it can be solved by the DMFT solver,
which can be for example Quantum Monte Carlo (QMC),
exact diagonalization, or a number of approximate methods, such as SPTF\cite{katsnelson02epjb30:9}.
The DMFT solver is the central part of the LDA+DMFT scheme.
It uses the bath GF $\lk\mathcal{G}_0\rk_{\xi_1, \xi_2}(z)$
and produces the new self energy $\Sigma_{\xi_1, \xi_2}(z)$.
The equations (\ref{dcount}), (\ref{gloc}), (\ref{gf}), (\ref{dmftsigma}), (\ref{g0def}) and (\ref{action})
constitute the DMFT cycle which is solved self-consistently until the convergence is reached.
The number of electrons is given by
\be
N = \lim_{\delta \to +0} T \sum_{i \omega_n} e^{i \omega_n \delta} \Tr  G(i \omega_n) =
T \sum_{i \omega_n} \Tr \lk  G(i \omega_n) + \frac 12\rk.
\ee
This equation is used to determine the LDA+DMFT chemical potential (Fermi energy), which must produce
the correct number of electrons.
In the following subsection we show how the DMFT equations can be presented in a given LDA basis set.

\subsection{DMFT equations for a given LDA basis set}
The Kohn-Sham eigenfunctions belong to the Hilbert space of one-electron states in
a solid. The choice of the basis set in this space is dictated by the method (LMTO, LAPW, \ldots),
and we can use a basis either in the real-space or in the reciprocal-space.
The real space basis set $\lfi \ket{\vvR, \chi} \rfi$ is defined by the wavefunction
\be
\psi_{\vvR, \chi} (\vvr)  \equiv \psi_{\chi} (\vvr - \vvR)
\ee
which is typically localized in a small area around the lattice site $\vvR$.
The k-space basis set $\lfi \ket{\vk, \chi} \rfi$ is a basis set that satisfies the Bloch theorem:
for any translation vector $\vT$
\be
\psi_{\vk, \chi} (\vvr + \vT) = e^{i \vk \vT} \psi_{\vk, \chi} (\vvr),
\ee
where $\vk$ belongs to the Brillouin zone.
There is a one-to-one correspondence between real-space and $\vk$-space basis sets,
given by the Fourier transformation
\be
\label{rs-four}
\ket{\vk, \chi} = \sum_{\vvR} e^{i \vk \vvR} \ket{\vvR, \chi},
\ket{\vvR, \chi} = \sum_{\vk} e^{ -i \vk \vvR} \ket{\vk, \chi},
\ee
where
\be
\sum_{\vk} \equiv \frac 1{V_{BZ}} \int_{BZ} \mathrm{d}\vk , \quad \sum_{\vk} 1 = 1.
\ee

Since the basis set $\lfi \ket{\vvR, \chi} \rfi$ or $\lfi \ket{\vk, \chi} \rfi$ in general is not orthogonal and not normalized,
the linear algebra becomes more cumbersome.
The overlap matrix is
\be
S_{\chi_1, \chi_2} = \braket{\chi_1| \chi_2}.
\ee
The conjugate basis set $\lfi \ket{\tilde\chi} \rfi$ is defined by the relations
\be
\label{conjg}
\braket {\tilde\chi_1|\chi_2} = \braket {\chi_1|\tilde\chi_2} = \delta_{\chi_1, \chi_2}, \quad \sum_{\chi} \ket{\tilde\chi}\bra{\chi} = \hat 1,
\ee
or, explicitly,
\be
\label{conjg2}
\ket{\tilde\chi_1} = \lk S^{-1} \rk_{\chi_2, \chi_1} \ket{\tilde\chi_2}, \quad
\bra{\tilde\chi_1} = \lk S^{-1} \rk_{\chi_1, \chi_2} \bra{\tilde\chi_2}.
\ee
If, and only if, the basis set $\lfi \ket{\chi} \rfi$ is orthogonal and normalized,
then $\lfi \ket{\tilde\chi} \rfi$ coincides with $\lfi \ket{\chi} \rfi$.

We use the following definition for the matrix elements of an operator
\be
A_{\chi_1, \chi_2} = \braket{\chi_1 | \hA |\chi_2}, \quad \hA = \sum_{\chi_1, \chi_2} \ket{\tilde\chi_1} A_{\chi_1, \chi_2} \bra{\tilde\chi_2},
\ee
and in this subsection we always put a hat above an operator to distinguish operators from matrices.
This convention leads to the following rules of operator-to-matrix correspondence
\begin{eqnarray}
\hA  & \to  & A  \qquad  \qquad \text{operator} \\
\hat 1 & \to &  S  \qquad  \qquad \text{unity operator} \\
\hA \hB & \to &  A S^{-1} B   \qquad   \text{product of two operators} \\
\label{lainv} \hA^{-1} & \to & S A^{-1} S  \qquad \text{inverse of an operator}
\end{eqnarray}

The LDA $\vk$-space basis set $\lfi \ket{\vk, \chi} \rfi$ should be used in order to calculate
the local GF (\ref{gloc}), since we know the LDA Hamiltonian matrix $h_{LDA}(\vk)$ and
the overlap matrix $S(\vk)$ in this basis set. On the other hand, the DMFT impurity problem is formulated
for the correlated subspace with a real-space basis set
$\lfi \ket{\vvR, \xi} \rfi$. Transforming back and forth between the local GF's and the self energies is
thus necessary at each DMFT iteration.
Using Eqs. (\ref{conjg})--(\ref{lainv}), it is easy to show that Eq. (\ref{gloc}) becomes
\begin{widetext}
\be
\label{gloc2}
\lk G_{\vvR} \rk_{\xi_1, \xi_2} (z) = \sum_{\vk,\chi_1,\chi_2} \braket{\xi_1|\vk,\chi_1}
\lkv S(\vk) (z - \mu) - h_{LDA}(\vk) - \Sigma(\vk,z) \rkv^{-1}_{\chi_1, \chi_2}  \braket{\vk,\chi_2 | \xi_2},
\ee
\end{widetext}
where $\Sigma(\vk,z)$ is the self energy matrix in the LDA basis $\lfi \ket{\vk, \chi} \rfi$.
This expression would be exact only if the basis set $\lfi \ket{\vk, \chi} \rfi$ was complete.
It would also be exact if the correlated orbitals $\ket{\vvR, \xi}$ belonged to the space spanned
by the basis functions $\ket{\vk, \chi}$, like for the  orthogonalized
MTO's, since it would mean that $\lfi \ket{\vk, \chi} \rfi$ is complete within the space of interest.
In realistic full-potential calculations the completeness of $\lfi \ket{\vk, \chi} \rfi$
is a reasonable approximation. The local self energy transformed into
the LDA basis is in turn given by
\be
\label{sig2}
\Sigma_{\chi_1, \chi_2} (\vk,z) = \sum_{\xi_1,\xi_2} \braket{\vk,\chi_1|\xi_1} \Sigma_{\xi_1, \xi_2} (z)
\braket{\xi_2|\vk,\chi_2}.
\ee

The one-particle excitation spectrum of a system is given by the
density of states (DOS)
\be
\label{dos}
D(\epsilon) = - \frac 1{\pi} \Tr \lkv \imag \hat G(\epsilon + i 0) \rkv,
\ee
and by the spectral density, which is the $\vk$-resolved DOS
\be
\label{specf}
A(\vk,\epsilon) = - \frac 1{\pi} \sum_{\chi} \braket{\vk,\tilde\chi| \imag \hat G(\epsilon + i 0)  | \vk,\chi}.
\ee
The spectral density generalizes the concept of quasiparticle band structure by
allowing quasiparticles to decay, thus introducing smearing of bands.
In the absence of self energy it reduces to the usual Kohn-Sham band structure
\[
A_{KS}(\vk,\epsilon) = \sum_{n} \delta \lk \epsilon - \epsilon_n(\vk) \rk.
\]
As we already mentioned, typical spectral density has coherent (quasiparticles) features and also possibly non-coherent
(dispersionless) ones: Hubbard band satellites.
Note that DMFT gives Green function at Matsubara frequencies $i \omega_n$, while
DOS and the spectral density are defined via GF at the $\epsilon + i 0$ contour. The
numerical analytical continuation can be done, for example, using the Pade approximation.

\section{\label{ch:brianna}Implementation}

\begin{figure}
\includegraphics[scale=0.3]{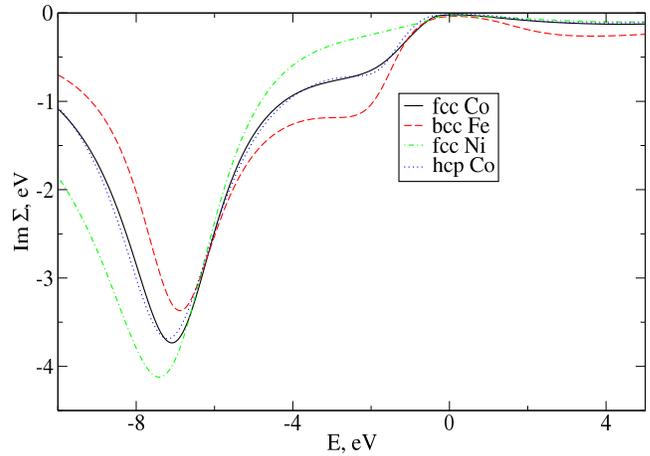}
\caption{\label{f:sig-up} (Color online) Imaginary part of the self energy for iron, cobalt and nickel, majority spin
(MT correlated orbitals at $T=400$ K).}
\end{figure}

\begin{figure}
\includegraphics[scale=0.3]{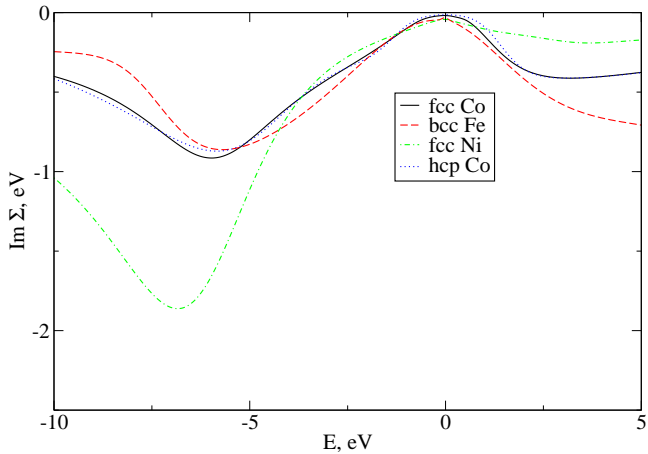}
\caption{\label{f:sig-dn} (Color online) Imaginary part of the self energy for iron, cobalt and nickel, minority spin
(MT correlated orbitals at $T=400$ K).}
\end{figure}

\begin{figure}
\includegraphics[scale=0.3]{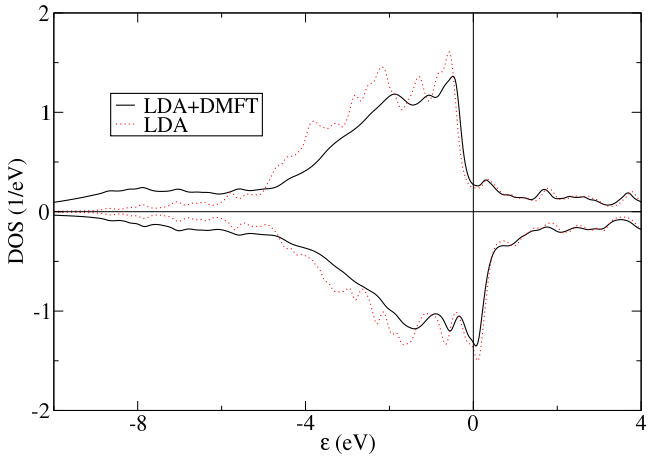}
\caption{\label{f:ni-dos} (Color online) Density of states of fcc nickel (LDA+DMFT with MT correlated orbitals at
$T=400$ K vs LDA).}
\end{figure}

\begin{figure*}
\includegraphics[scale=0.8]{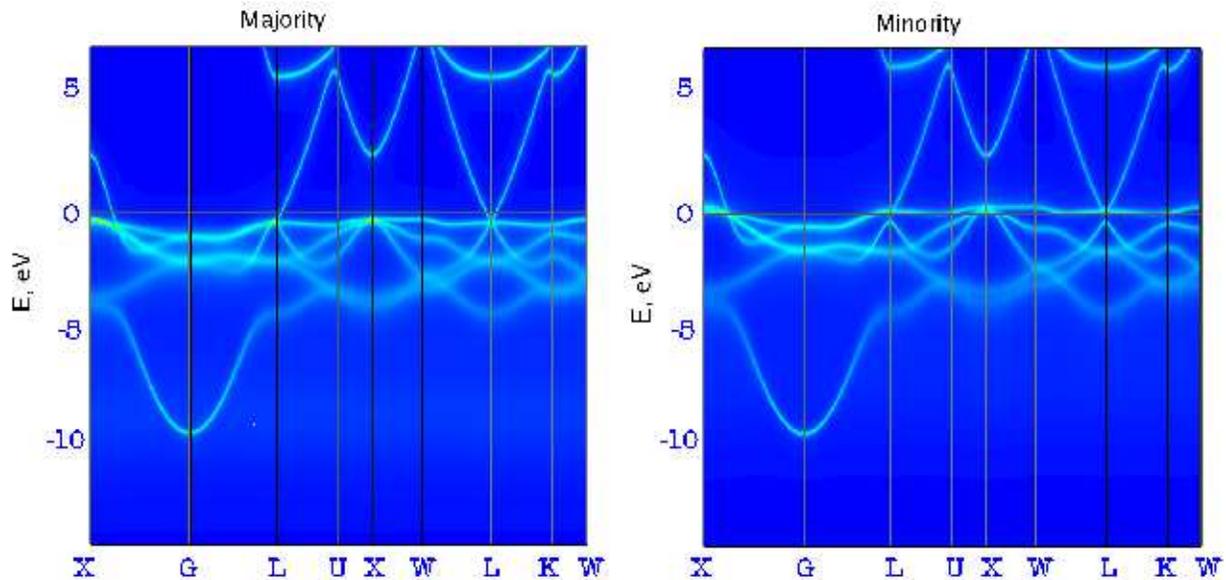}
\caption{\label{f:ni-spec} (Color online) Spectral density of fcc nickel from LDA+DMFT with MT correlated orbitals at
$T=400$ K.}
\end{figure*}

\begin{figure}
\includegraphics[scale=0.3]{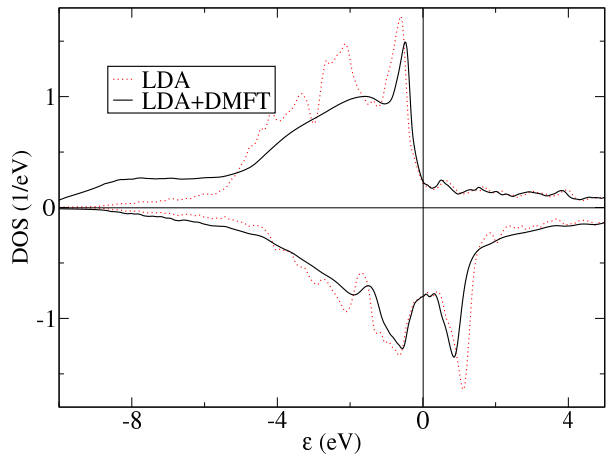}
\caption{\label{f:hcpco-dos} (Color online) Density of states of hcp cobalt (LDA+DMFT with MT correlated orbitals at
$T=400$ K vs LDA)}
\end{figure}

\begin{figure*}
\includegraphics[scale=0.8]{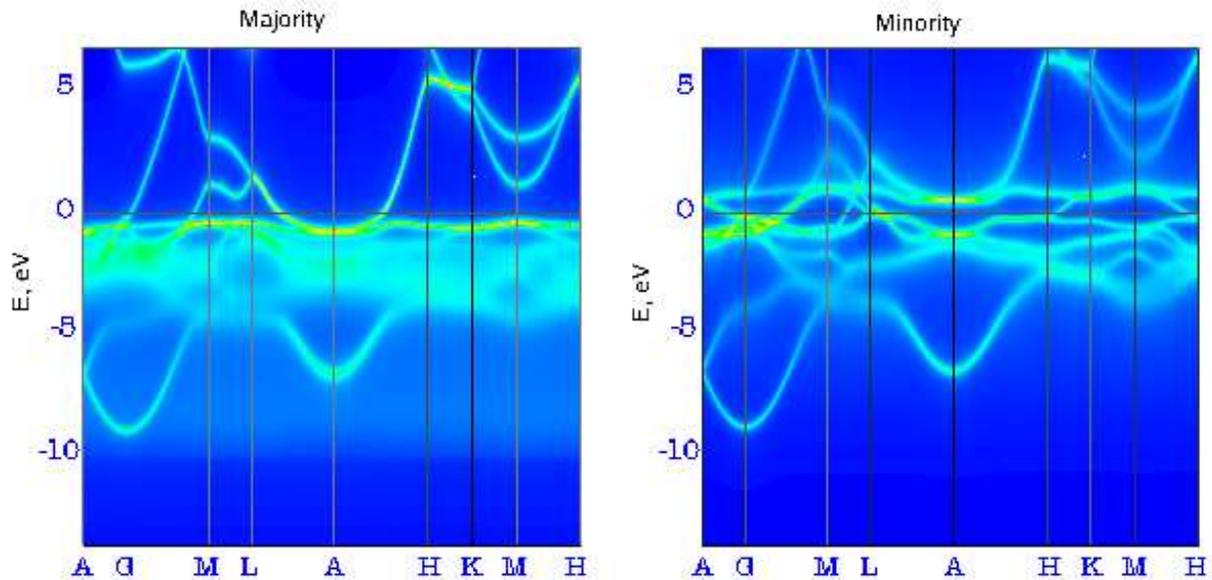}
\caption{\label{f:hcpco-spec} (Color online) Spectral density of hcp cobalt from LDA+DMFT with MT correlated orbitals at
$T=400$ K.}
\end{figure*}

\begin{figure}
\includegraphics[scale=0.3]{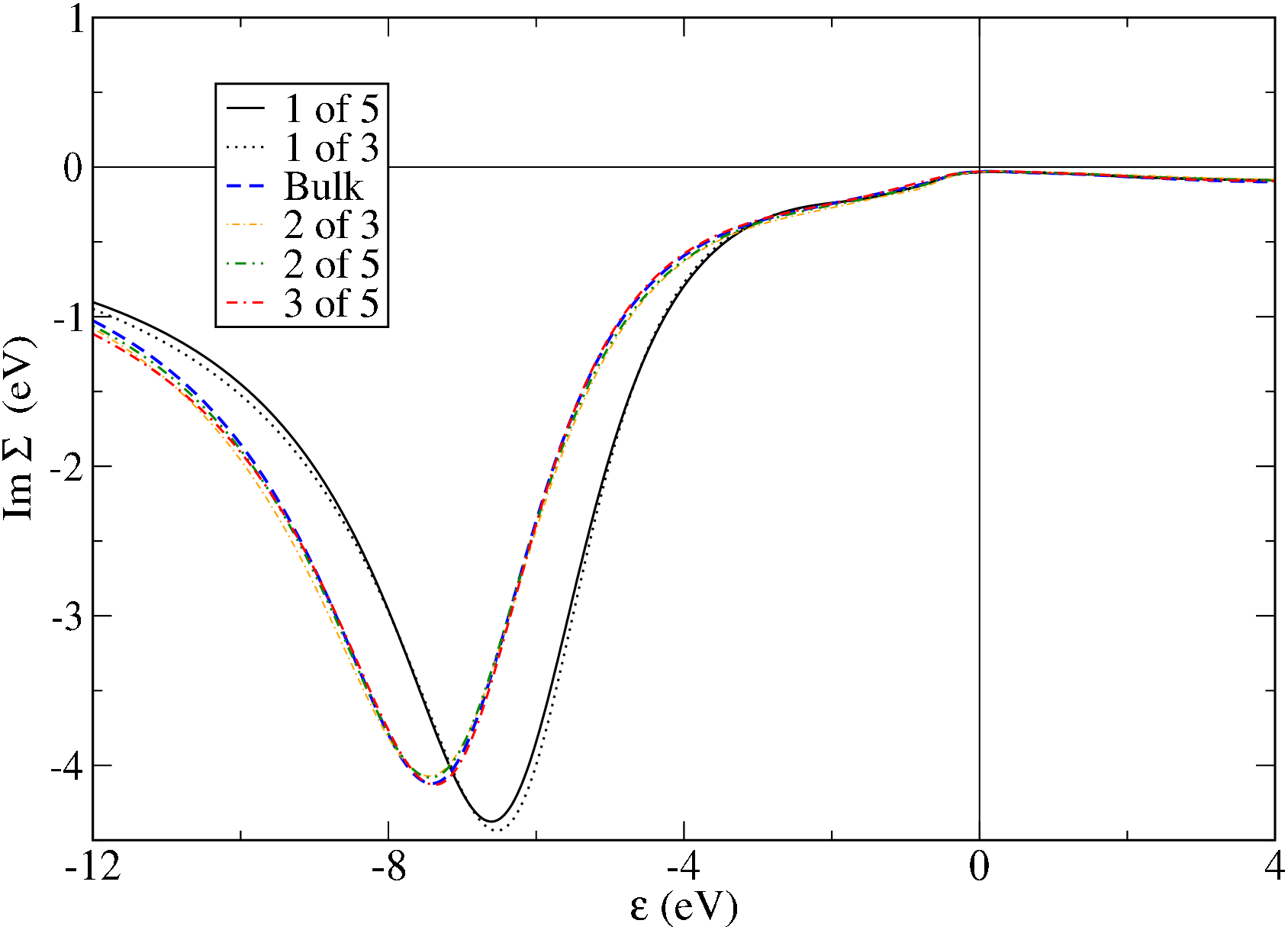}
\caption{\label{f:ni5-sig-up} (Color online) Imaginary part of the self energy for nickel
5-layer (111) slab, 3-layer (111) slab and bulk fcc nickel, majority spin (from LDA+DMFT with MT correlated orbitals at
$T=400$ K). In the legend "1 of 5" indicates
 the surface atom of the 5-layer slab, "2 of 5" the sub-surface atom and "3 of 5" the quasi-bulk atom. Similarly "1 of
 3" indicates the surface atom of the 3-layer slab and "2 of 3" the quasi-bulk atom.}
\end{figure}

\begin{figure}
\includegraphics[scale=0.3]{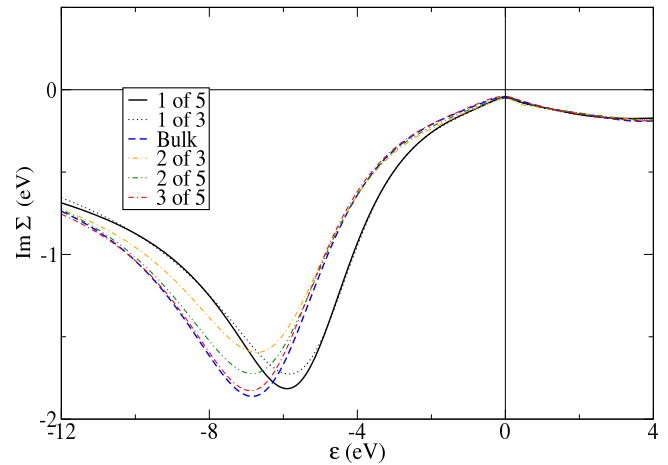}
\caption{\label{f:ni5-sig-dn} (Color online) Imaginary part of the self energy for nickel 5-layer and 3-layer (111)
slabs and bulk, minority spin (from LDA+DMFT with MT correlated orbitals at $T=400$ K). In the legend "1 of 5" indicates
 the surface atom of the 5-layer slab, "2 of 5" the sub-surface atom and "3 of 5" the quasi-bulk atom. Similarly "1 of
 3" indicates the surface atom of the 3-layer slab and "2 of 3" the quasi-bulk atom.}
\end{figure}

\begin{figure*}
\includegraphics[scale=0.8]{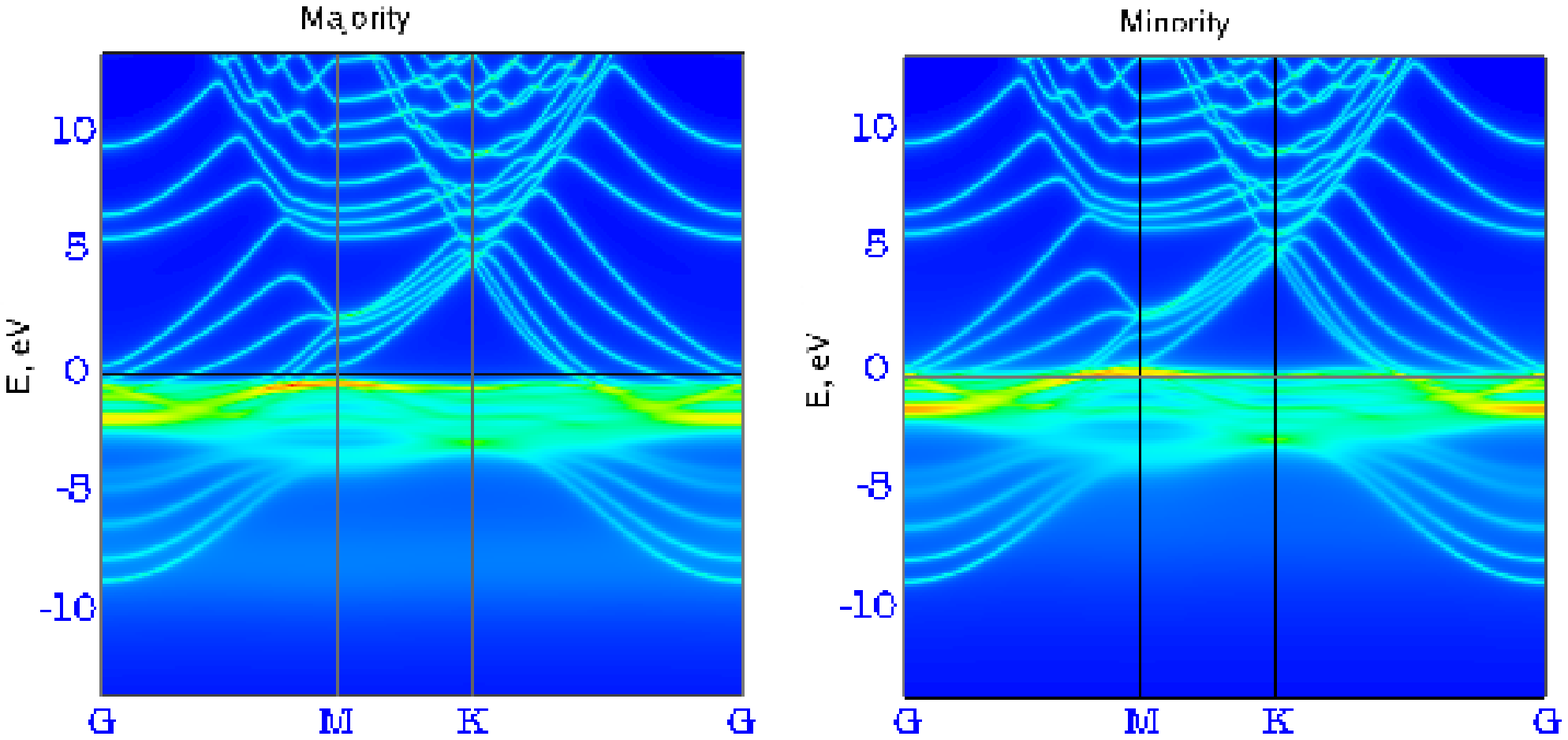}
\caption{\label{f:ni5-spec} (Color online) Spectral density of nickel (111) surface modelled by the 5-layer slab (from
LDA+DMFT with MT correlated orbitals at $T=400$ K).}
\end{figure*}

In this paper we introduce the code BRIANNA, a new LDA+DMFT
implementation based on the full-potential linear muffin tin
orbital (FP-LMTO) code developed in
Ref.\onlinecite{wills:fp-lmto}. As we already mentioned, FP-LMTO
gives an accurate description of solids within LDA, and the
full-potential treatment is especially important for open
structures and surfaces. On the other hand, FP-LMTO uses a
relatively small basis set, which is convenient for calculating
Green functions, since it involves inverting a matrix in the LDA
basis set for each Matsubara frequency and $\vk$-point. The
typical basis set is ``double-minimal'', i.e. it contains two
basis functions per each site and $l,m,\sigma$, with different
``tail energies''. It is still much smaller than the basis set in
the plane wave and augmented plane wave based codes.

\begin{figure}
\includegraphics[scale=0.3]{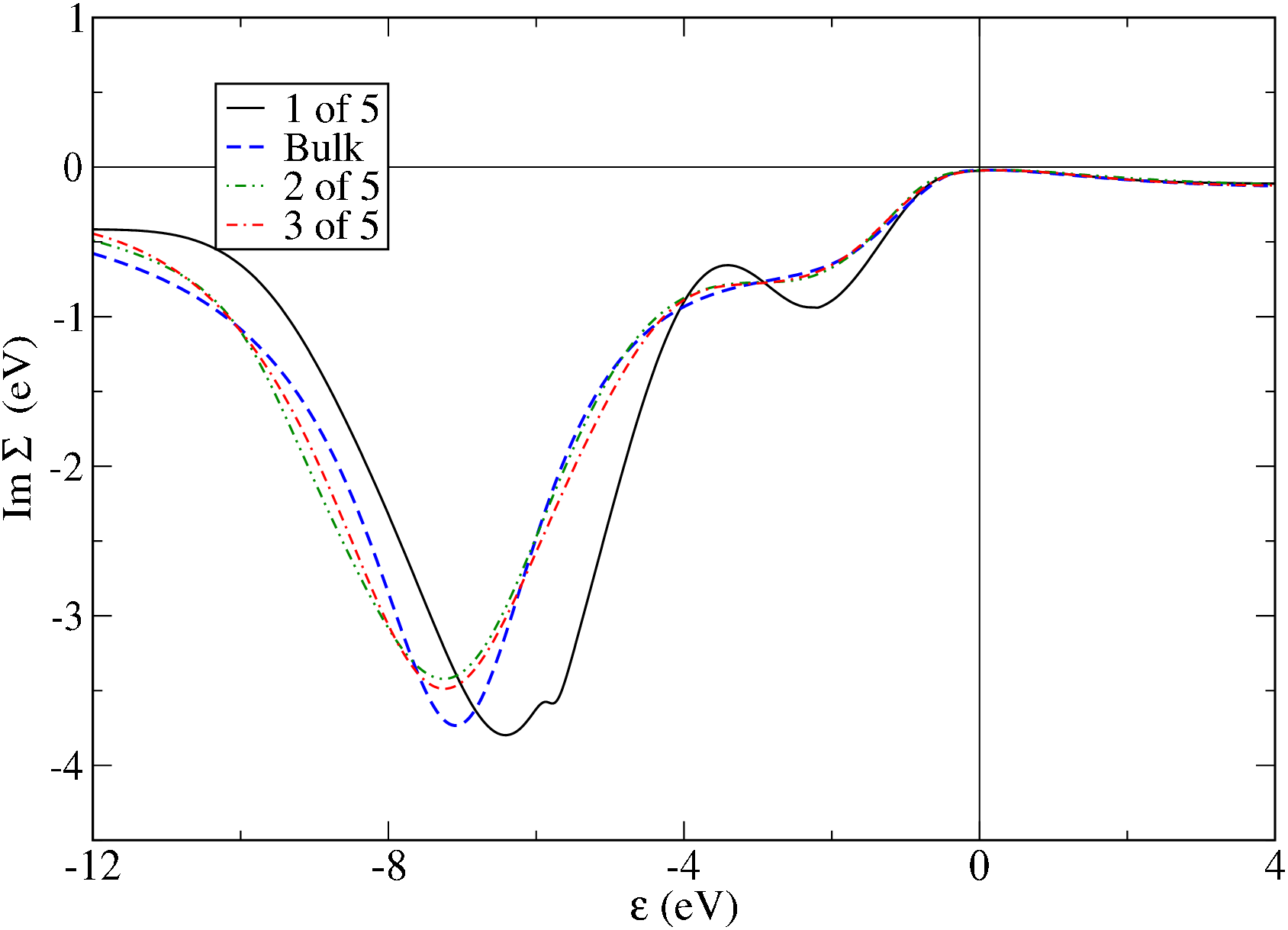}
\caption{\label{f:co5-sig-up} (Color online) Imaginary part of the self energy for fcc cobalt 5-layer (111) slab,
majority spin (from LDA+DMFT with MT correlated orbitals at $T=400K$ K). In the legend "1 of 5" indicates the surface
 atom of the slab, "2 of 5" the sub-surface atom and "3 of 5"
the quasi-bulk atom.}
\end{figure}

\begin{figure}
\includegraphics[scale=0.3]{fig16.eps}
\caption{\label{f:co5-sig-dn} (Color online) Imaginary part of the self energy for fcc cobalt 5-layer (111) slab,
minority spin (from LDA+DMFT with MT correlated orbitals at $T=400K$ K). In the legend "1 of 5" indicates the surface
 atom of the slab, "2 of 5" the sub-surface atom and "3 of 5"
the quasi-bulk atom.}
\end{figure}

\begin{figure*}
\includegraphics[scale=0.8]{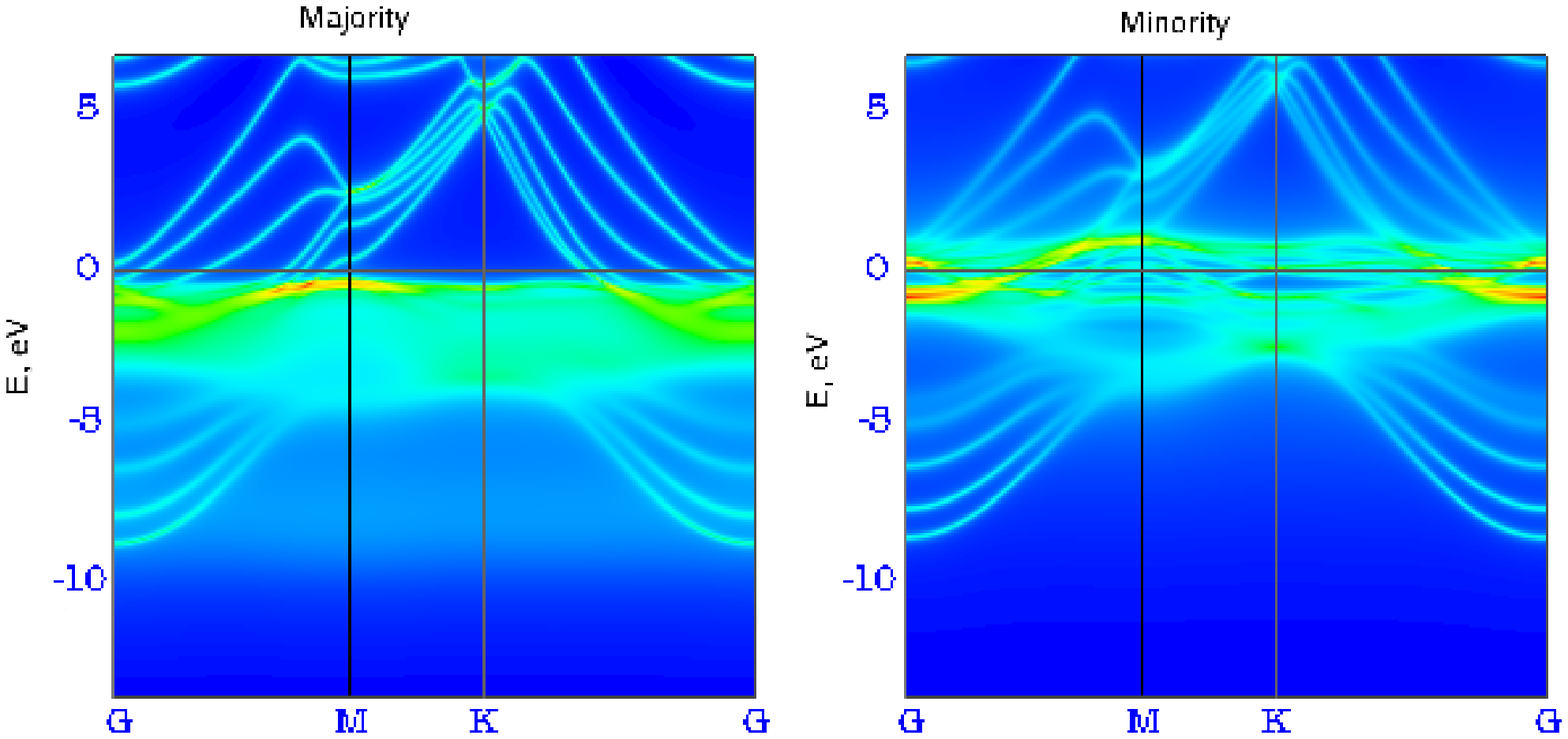}
\caption{\label{f:co5-spec} (Color online) Spectral density of fcc cobalt (111) surface modelled by the 5-layer slab
(from LDA+DMFT with MT correlated orbitals at $T=400$ K).}
\end{figure*}

\begin{figure}
\includegraphics[scale=0.3]{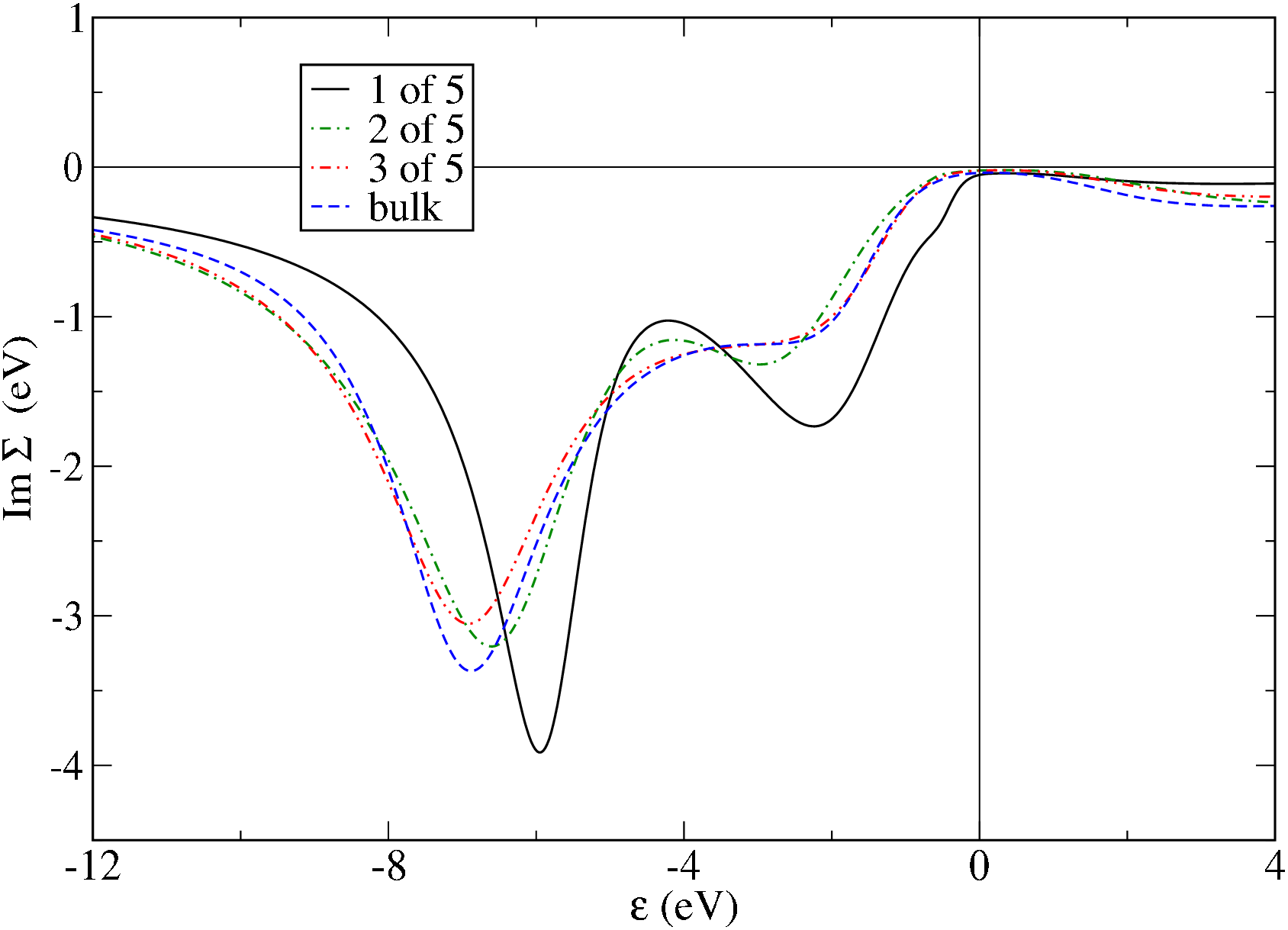}
\caption{\label{f:fe5-sig-up} (Color online) Imaginary part of the self energy for bcc iron (001) surface
 modelled by the 5-layer slab, majority spin (from LDA+DMFT with MT correlated orbitals at $T=400$ K).  In the legend
  "1 of 5" indicates the surface atom of the slab, "2 of 5" the sub-surface atom and "3 of 5"
the quasi-bulk atom.}
\end{figure}

\begin{figure}
\includegraphics[scale=0.3]{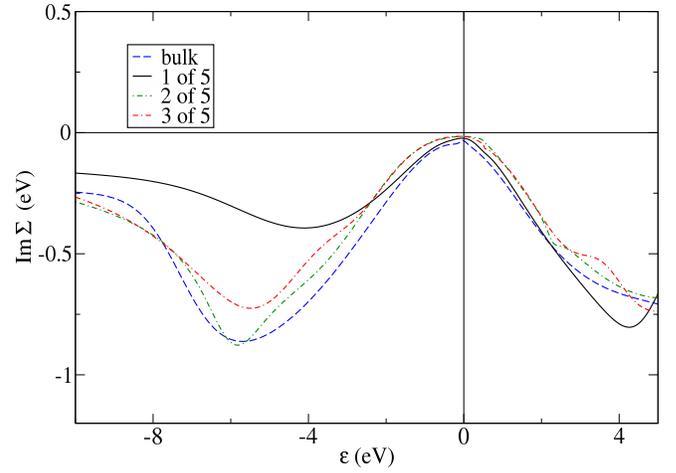}
\caption{\label{f:fe5-sig-dn} (Color online) Imaginary part of the self energy for bcc iron (001) surface,
modelled by the 5-layer slab, minority
spin (from LDA+DMFT with MT correlated orbitals at $T=400$ K).  In the legend "1 of 5" indicates the surface atom
 of the slab, "2 of 5" the sub-surface atom and "3 of 5" the quasi-bulk atom.}
\end{figure}

\begin{figure*}
\includegraphics[scale=0.8]{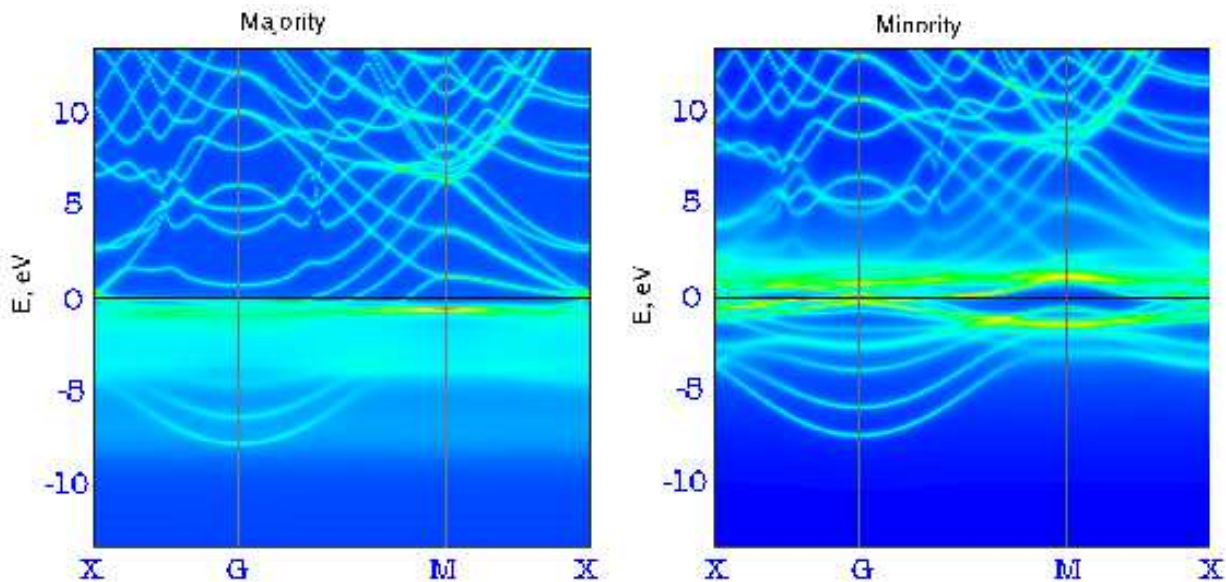}
\caption{\label{f:fe5-spec} (Color online) Spectral density of bcc iron (001) surface modelled by the 5-layer slab (from
LDA+DMFT with MT correlated orbitals at $T=400$ K).}
\end{figure*}

\begin{figure}
\includegraphics[scale=0.3]{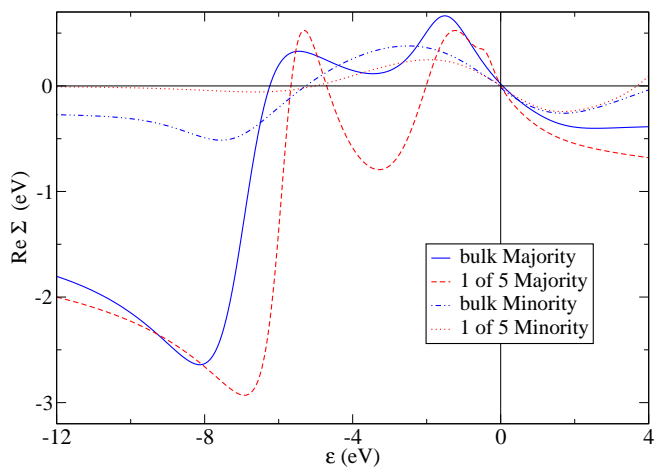}
\caption{\label{f:Fe5-sig-re-com-bulk} (Color online) Real part of the self energy for bcc iron (001) surface
 modelled by the 5-layer
 slab. Both majority and minority spins  for the surface atom (labelled "1 of 5") are reported and 
 compared to the bulk values (from LDA+DMFT with MT correlated orbitals at $T=400$ K).}
\end{figure}

\begin{figure*}
\includegraphics[scale=0.8]{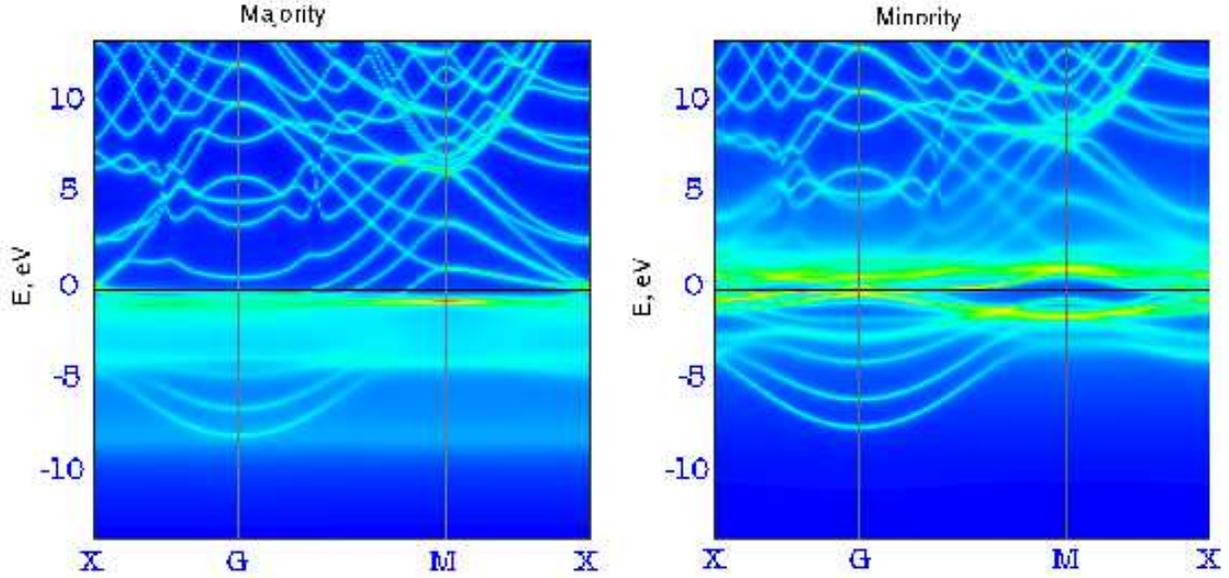}
\caption{\label{f:fe5_spec_manyU} (Color online) Spectral density of bcc iron (001) surface modelled by the 5-layer
 slab and
with different values of $U$ for atoms belonging to different layers of the slab (from 
LDA+DMFT with MT correlated orbitals at $T=400$ K).}
\end{figure*}

\begin{figure}
\includegraphics[scale=0.3]{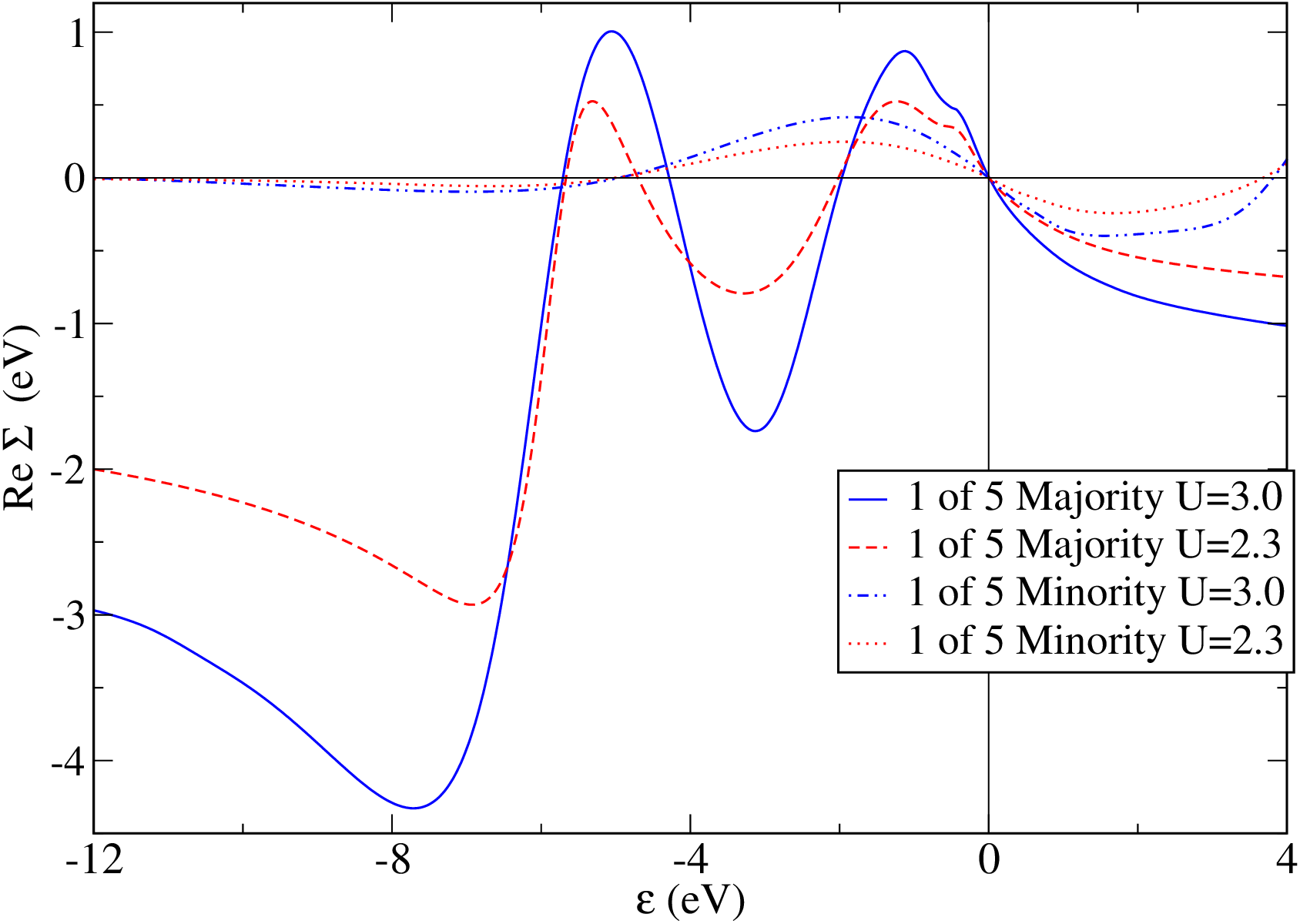}
\caption{\label{f:Fe5-sig-re-com-manyU} (Color online) Real part of the self energy for bcc iron (001) surface
 modelled by the 5-layer
 slab and with different values of $U$ for atoms belonging to different layers of the slab. Both majority and minority 
 spins for the surface atom (labelled "1 of 5") are reported and compared to the
 values calculated in the simulation with constant $U$ (from LDA+DMFT with MT correlated orbitals at $T=400$ K).}
\end{figure}

\begin{figure}
\includegraphics[scale=0.3]{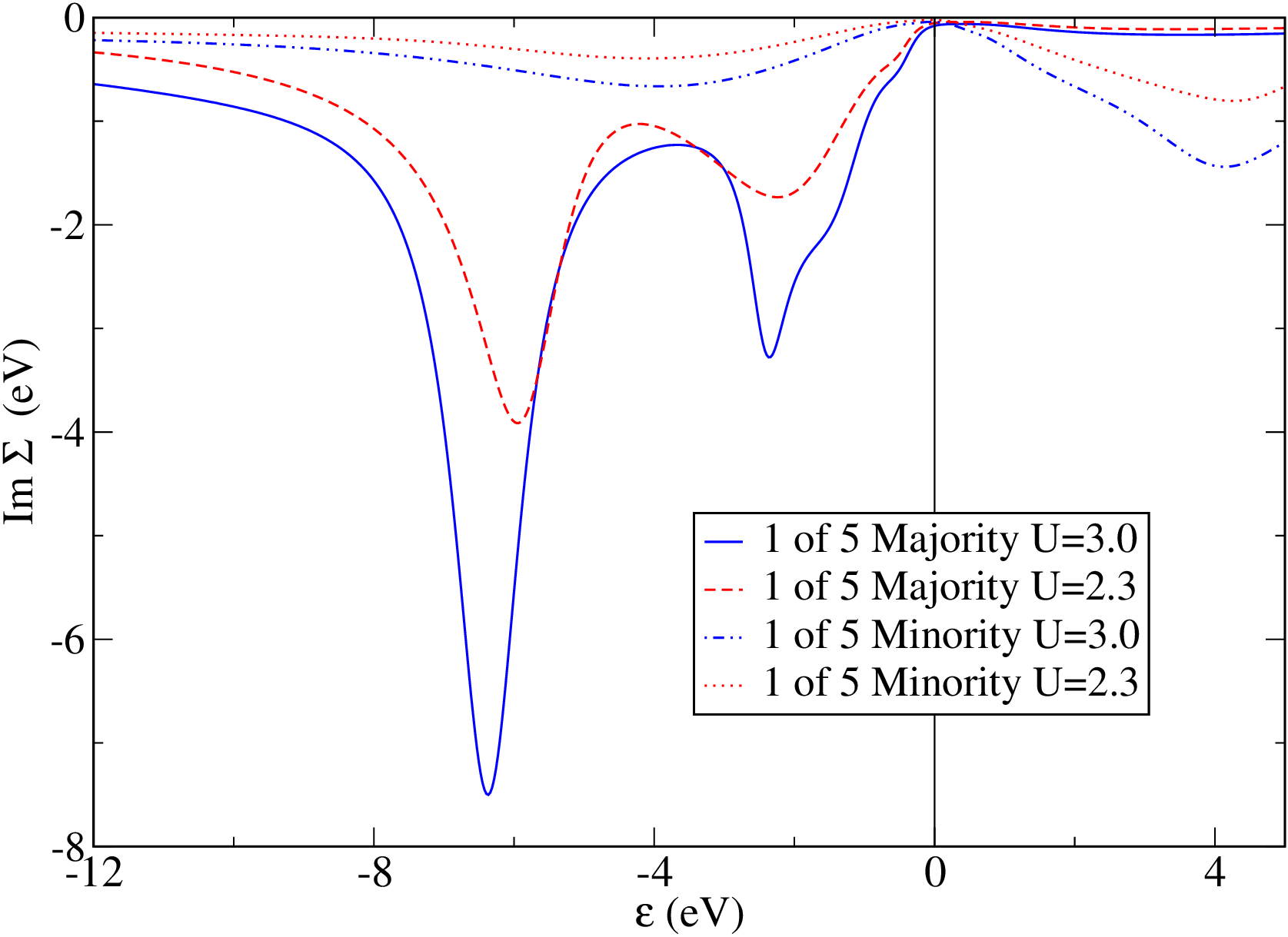}
\caption{\label{f:Fe5-sig-im-com-manyU} (Color online) Real part of the self energy for bcc iron (001) surface
 modelled by the 5-layer slab and with different values of $U$ for atoms belonging to different layers of the slab.
 Both majority and minority  spins for the surface atom (labelled "1 of 5") are reported and compared to the
 values calculated in the simulation with constant $U$ (from LDA+DMFT with MT correlated orbitals at $T=400$ K).}
\end{figure}

There are a few other technical issues worth mentioning here. A DMFT solver typically
needs about 1000 or more Matsubara frequencies (above the real axis). Instead of performing 
matrix inversion in Eq. (\ref{gloc2})
for every frequency (and every $\vk$-point), it is a standard technique nowadays
to use a smaller (usually logarithmic) mesh in Eq. (\ref{gloc2}) and, after having calculated the local GF,
transform it to the Matsubara mesh using cubic splines. Inverse transformation is applied to
the self energy in order to plug it into Eq. (\ref{gloc2}). All calculations of the present paper
use 1024 Matsubara frequencies in the DMFT solver, but only 80 points in Eq. (\ref{gloc2}). The first
16 of them coincide with the first Matsubara frequencies, while the rest forms a logarithmic mesh.
Another issue involves numerical analytical continuation in (\ref{dos}) and (\ref{specf}).
Analytical continuation using the Pade approximation can introduce serious numerical errors.
In the present implementation we do not apply the Pade approximation to the Green functions.
Instead, we use it only for the self energy $\Sigma(z)$,
while the GF is calculated directly at the $\epsilon + i 0$ contour using Eq. (\ref{gf}).

We have already mentioned the very important question of choosing
the orbitals spanning the correlated subspace. In this paper we
use two different definitions, both of them atomic-like and
derived from transition metal d states. The first, more
traditional, uses orthogonalized d-type basis functions of the
FP-LMTO method, transformed to the real space via (\ref{rs-four}). We
call this definition orthogonalized LMTO (ORT) correlated
subspace. We remind the reader that it is poorly localized, and also that the
orbitals $\ket{\vvR, \xi}$ do not have pure $l,m$ character. Such
kind of
approach requires minimal LDA basis set for the d-type electrons.
This FP-LMTO code\cite{wills:fp-lmto} allows to use a
double-minimal basis set (two or even more energy tails) for
sp-electrons and a minimal one (single tail) for d electrons. The
latter have less dispersion than sp electrons, and use of the
single-tail basis set for them does not lead to any severe errors
within LDA.

Our second choice is somewhat opposite, since it deals with extremely localized correlated orbitals.
We call it muffin-tin only (MT) correlated subspace. $\ket{\vvR,\xi}$ is chosen as
\be
\label{xi-def}
\Psi_{\vvR,\xi}(\vvr) = \begin{cases}
 \Phi_l(\lab \vvr - \vvR_{\xi} \rab) Y_{lm} (\widehat{\vvr - \vvR_{\xi}}) , &  \lab \vvr - \vvR_{\xi} \rab  < R_{\text{MT},\xi} \\
 0 , &  \lab \vvr - \vvR_{\xi} \rab  > R_{\text{MT},\xi}
\end{cases},
\ee
$\vvR_{\xi} \equiv \vvR + \vvr_{\xi}$ is the site where the orbital $\ket{\vvR,\xi}$ is located
and $R_{\text{MT},\xi}$ is the muffin-tin radius for this site.
The pure muffin-tin radial function $\Phi_l(r)$ is 
the solution of the radial Schr\"odinger equation in the spherically averaged
Kohn-Sham potential\cite{wills:fp-lmto}, inside the muffin-tin only, for a
certain energy $E_{\nu}$.

The correlated orbital in Eq. (\ref{xi-def}) is zero outside a given muffin-tin,
and is thus ultimately local. The correlated orbitals have pure angular momentum character,
but, at the same time, they are orthogonal by definition (since they do not overlap).
Note that the correlated orbitals obviously do not form a
complete basis set within the Hilbert space of one-electron
wafefunctions (since the interstitial region is not included at all).
This is not a problem, since they
are only used to define the Hubbard-U term
in the Hamiltonian (\ref{hlpu}), while the "hopping" term is the LDA Hamiltonian
defined using the FP-LMTO basis set, which we assume to be sufficiently complete.

\section{\label{ch:results}Results}
\subsection{Iron}

The LDA+DMFT self energies of bcc iron have been calculated for $U = 2.3$ eV, $J = 0.9$ eV.
We used both muffin-tin only (MT) and the orthogonalized LMTO (ORT) sets of correlated orbitals, and also
performed calculations for different temperatures.
In Figures \ref{f:fe-sig-up}  and \ref{f:fe-sig-dn} we present the imaginary part of the self energies,
$\imag\Sigma(\epsilon + i 0)$, for majority and minority spins respectively, while
in Figure \ref{f:fe-sig-re} we show the real part of $\Sigma$ (only for the MT basis and for $T=1500$ K).
In quasiparticle language the matrix elements $\rere\Sigma(E+i0)_{\xi,\xi}$ describe the shift of the
quasiparticle bands, while  $-\imag\Sigma(\epsilon+i0)_{\xi,\xi}$ has the physical meaning
of quasiparticle band smearing $\Gamma$, which is the inverse of the quasiparticle decay time $\tau = 1/\Gamma$.
The band structure is only well defined if $\left| \imag \Sigma_{\xi,\xi} \right| \ll W$, where $W$ is the bandwidth.
For metals this is always true in the vicinity of the Fermi level, since $\imag\Sigma(+i0) = 0$.
The self energies presented here are averaged over the orbital indices $m$, namely
\be
\Sigma(z) \equiv \frac 15 \sum_{m} \Sigma_{mm}(z).
\ee
The crystal field splitting of $\Sigma$ is rather small and we are not going to discuss it in details.
Figure \ref{f:fe-dos} shows the total density of states (DOS) of bcc iron (LDA+DMFT vs LDA), while
the spectral density ($\vk$-resolved DOS) is presented in Figure \ref{f:fe-spec} for
several high-symmetry directions. We remind the reader that the spectral
density is the generalization of the band structure with finite quasiparticles lifetime taken into account.
Both figures use the MT basis set and $T=400$ K.

The three curves in Figs. \ref{f:fe-sig-up} and \ref{f:fe-sig-dn} are qualitatively similar,
proving that both MT and ORT correlated orbitals (corresponding to well-localized and poorly localized d-states,
respectively)
can be used to adequately describe iron within LDA+DMFT. However, the exact amplitude of the peaks in $\Sigma$
is sensitive to the choice of the correlated subspace. We are going to use the muffin-tin only (MT) correlated orbitals for the
rest of this paper. Note also that $\Sigma$ is practically temperature-independent
for a wide range of temperatures.

The majority spin $\imag \Sigma(\epsilon + i0)$ in Fig. \ref{f:fe-sig-up} has the main peak
at $\epsilon \simeq -7$ eV, by reaching the value $-3.4$ eV.
This gives rather strong damping of quasiparticles, as we can observe in Fig. \ref{f:fe-spec}.
There is also a shoulder or small minimum at $\epsilon \simeq -2$ eV.
The correlation effects are more pronounced for the majority spin electrons,
which is common for late transition metals (see Ref. \onlinecite{monastra02prl88:236402}
for an interesting discussion).
The LDA+DMFT density of states (Fig. \ref{f:fe-dos}) shows the narrowing of the majority-spin d-band
compared to the LDA DOS and also a satellite at $\epsilon \simeq -7$ eV. This is the effect
of $\rere \Sigma(\epsilon + i0)$. The positive region of $\rere \Sigma$ for the majority electrons
between -6 eV and the Fermi level in Fig. \ref{f:fe-sig-re} leads to the narrowing of the band,
while the sharp negative peak at -8 eV "draws" the electrons down in energy, leading to the formation of
the DOS satellite. Naturally,
the smearing of the quasiparticle bands, given by $\imag \Sigma$, leads to the smearing of
the sharp peaks of the LDA DOS. Note that our self energies and DOS differ somewhat
from the ones in Ref. \onlinecite{katsnelson99jpcm11:1037}. In particular, we clearly observe a DOS satellite
at $\epsilon \simeq -7$ eV, which was not observed in the earlier calculation for $U=2.3$ eV and $J=0.9$ eV,
but only for much larger values of $U$.
The reason, we believe, is that Ref. \onlinecite{katsnelson99jpcm11:1037} used a simplified
version of the SPTF solver, while in the present paper the full implementation of the
SPTF\cite{katsnelson02epjb30:9} is used. To the best of our knowledge this is the first calculation that shows the
existence of such a satellite, and this could open a new scientific problem, since it has never been
 reported in any experiment.

\subsection{Cobalt and nickel}

The LDA+DMFT self energies for fcc nickel, hcp cobalt and fcc cobalt are presented in Figs. \ref{f:sig-up}
and \ref{f:sig-dn}, and the bcc iron self energy is also shown for comparison. The values of Hubbard parameters were $U=2.3$ eV, $J=0.9$ eV
for cobalt and $U=3$ eV, $J=1$ eV for nickel. Strictly speaking the parameters $U$ and $J$ are somewhat arbitrary
(since they apply to a model LDA+U Hamiltonian) and their values depend on the choice of the correlated orbitals.
We make a rather traditional choice of $U$ and $J$\cite{monastra02prl88:236402,katsnelson99jpcm11:1037} in the present paper,
however\cite{katsnelson99jpcm11:1037,lichtenstein01prl87:067205,katsnelson02epjb30:9,monastra02prl88:236402}.

The general structure of the self energy is similar for Fe, Co and Ni. The correlation effects for majority spin
electrons are strongest for nickel and weakest for iron, at least for the values of $U$ and $J$ used here.
The shoulder at $-2$ eV is most pronounced for iron and practically disappears for nickel.
The self energy curves for fcc cobalt and hcp cobalt are almost identical. The correlation
for minority spin electrons (Fig. \ref{f:sig-dn}) are by far strongest in nickel, which has the lowest magnetic moment
of the three elements,
therefore the difference between majority and minority spin behavior is less profound in nickel
 compared to iron and cobalt.

Figures \ref{f:ni-dos} and \ref{f:hcpco-dos} present density of states of fcc nickel and hcp cobalt respectively,
while Figures \ref{f:ni-spec} and  \ref{f:hcpco-spec} show the spectral density for these materials.
The density of states for nickel is in a good agreement with previous LDA+DMFT calculations\cite{katsnelson02epjb30:9}.
Note that the SPTF solver places the majority-spin satellite at about $-7.5$ eV, while in experiment it
is observed at $-6$ eV. The spectral density of fcc nickel is, to the best of our knowledge, presented here
for the first time. Since bcc iron, fcc cobalt and fcc nickel have different crystal structure, their band
structures naturally look different. However, Fe, Co and Ni all have strong smearing
of majority-spin bands at about $-7$ eV dictated by the peak in the self energy (Fig. \ref{f:sig-up}),
and show a DOS satellite at about $-7.5$ eV.

The LDA+DMFT values of the spin magnetic moments are substantially equal to the LDA values (e.g. for bcc Fe
 we have $\mu=2.23 \mu_B$ per atom from the DMFT calculation which should be compared to $\mu=2.22 \mu_B$ per atom from
 LSDA, and for hcp Co we obtain $\mu = 1.54 \mu_B$ per atom from the DMFT calculation which should be compared to 
  $\mu = 1.57 \mu_B$ from LSDA). Indeed the problem of the effect of the correlations on the spin and orbital
magnetic moments is very interesting and will be the subject of further investigations in the near future.
\subsection{Surfaces}

We model the nickel (111) surface with slabs having different
number of close-packed atomic layers. Such slabs form a
superlattice with a 30 {\AA} thick layer of vacuum separating
them, with each slab having two (111) surfaces. The calculations
have been done for 5-layer slabs; however just for methodical aims, to
show the sensitivity to the computational details the results for
3-layer slabs are also presented on some figures. In Figs.
\ref{f:ni5-sig-up} and \ref{f:ni5-sig-dn} we present the LDA+DMFT
self energies (imaginary part) for nickel slabs for majority and
minority spin respectively. Data for each layer of the 3-layer and
5-layer slabs and for the bulk fcc nickel (for comparison) are
presented. Notice that the self energy
of a nickel atom at the surface is obviously quite different from the self energies for the rest of the atoms in the
 slab. The most noticeable effect is that the positions of the peaks are shifted and that the correlation effects
 for majority spin electrons seem to be more enhanced at the surface compared to bulk. We will encounter similar
 effects for the other surfaces studied here (see below). This shows that the effect of the correlations is different
 for the topmost surface layer, compared to the rest of the surface layers, and that the sub-surface layer already
 seems really bulk-like. This finding is an observation that is worthy experimental attention. The reasons
 of such a difference are rather obvious: due to the reduced coordination number of the
 surface atoms the bands become narrower,  which makes correlation effects more important.
 In addition the screening of the electron-electron interaction is less
 effective for the surface atoms, and this increases the value of the Hubbard U. Although the basic mechanisms
 are easily
 identified for why correlation effects are more important at the surface, we provide here a quantitative
  measure of this effect.

The spectral density of the nickel 5-layer slab is presented in Fig. \ref{f:ni5-spec} along high-symmetry directions
of the two-dimensional Brillouin zone.  For well-defined quasiparticles, each
band of the bulk nickel splits into five bands for the 5-layer slab. Some of the bands are surface states,
while the rest joins into the bulk continuum when the number of atomic layers go to infinity. In Fig. \ref{f:ni5-spec},
it is already possible to observe the surface states (isolated bands) and the hint of the bulk continuum formation
(several bands that are very close to each other).

Similar results are obtained for the fcc cobalt (111) surface and for bcc iron (001) surface, whose self energies are
respectively shown in Figs. \ref{f:co5-sig-up}-\ref{f:co5-sig-dn} and in Figs. \ref{f:fe5-sig-up}-\ref{f:fe5-sig-dn}.
We can notice two main differences with respect to the results for the nickel: for the majority spin the shift of the
 peak  and the increase of its depth for the atoms on the surface are stronger, while for the minority spin
the correlation effects are, somewhat surprisingly, decreased (slightly for Co and strongly for Fe).
In Fig. \ref{f:Fe5-sig-re-com-bulk} we
 can observe the real part of the self energies for the atoms on the iron surface, compared to the bulk values.
 It is especially interesting that for the surface layer the self energy of minority spin states is considerably
 suppressed, which leads to the fact that the satellite at $-7.5$ eV is almost totally polarized and possesses majority
 spin charachter. In Figs. \ref{f:co5-spec} and \ref{f:fe5-spec} we show the spectral densities for the
 surfaces of Co and Fe, and
 the effects of the change of the imaginary parts of $\Sigma$ with respect to the bulk are clearly evident in the
difference of the definition of the quasi-particle bands for majority and minority spins.

 Finally we have to notice that the choice of the values of $U$, already non trivial for the bulk materials, becomes
 more problematic for the surfaces, where the screening is much smaller. To analyze this problem, we have tried to
 model the bcc Fe (001) surface with different values of $U$ for atoms belonging to different layers,
 namely $U=2.3$ for the inner layer, $U=2.4$ for the intermediate one and $U=3.0$ for the external one.
 In Figs. \ref{f:Fe5-sig-re-com-manyU} and \ref{f:Fe5-sig-im-com-manyU}
  we respectively show the real and the imaginary part of the self energy for the external atoms. In comparison to the
  previous calculation we do not observe any drastic effect, but mainly a reasonable increasing of the peaks and a
  small shift of the $-7.5$ eV satellite. This makes the satellite more pronounced in the density of states. The
  spectral densities are reported in Fig. \ref{f:fe5_spec_manyU}.

\section{Conclusion}

In this paper we have introduced the new full-potential LDA+DMFT code BRIANNA and
we have applied it to the correlated electronic structure of bulk Fe, Co, Ni, and the fcc Co and Ni (111) surface,
and the bcc Fe (001) surface.
The calculated self energies, DOS and the spectral densities are presented. The spectral density plots show
the corelated electronic structure in the most clear way, as the $\vk$-resolved DOS (or, equivalently, as  the
smeared band structure). The main correlation effects in iron, cobalt and nickel are observed for the
majority spin electrons and they include strong quasiparticle damping for at about $-7$ eV,
narrowing of the d-band (compared to LDA/GGA) and the appearance of a DOS satellite at about $-7.5$ eV, which
is a non-quasiparticle feature.

The calculations for Ni and Co (111) surfaces and for Fe (001)
surface show that the electron self energy depends mostly on the
local coordination number, with the atoms in the second layer from
the surface already being similar to the bulk. Hence our calculations
suggest that the effect of correlations should be different for the
surfaces of these elements, compared to the bulk. In addition, the spectral
density of the Ni (111) surface show both bulk and surface states.
The question ``How do the correlation effects depend on the
dimensionality of the problem?'' still needs further
investigation, however, and the the LDA+DMFT studies of slabs of
different thickness and nanowires are the subject of the future
research.
\\
\\

\section*{ACKNOWLEDGEMENTS}
We are grateful to L. V. Pourovskii for helpful discussions.
This work was supported by the Stichting voor Fundamenteel Onderzoek
der Materie (FOM), the Netherlands, and by the EU Research Training Network
``Ab-initio Computation of Electronic Properties of $f$-electron
Materials'' (contract: HPRN-CT-2002-00295). O.E. acknowledges support 
from the 
Swedish Research Council, The Foundation for Strategic Research, NSC  
and UPMAX,
and A. L. - support from the DFG (Grants  No. SFB 668-A3).

\bibliography{strings,kylie}

\end{document}